# Chiral control of electron transmission through molecules

Spiros S. Skourtis,[1] David N. Beratan,[2] Ron Naaman,[3] Abraham Nitzan[4] and David H. Waldeck[5]

[1]Department of Physics, University of Cyprus, Nicosia 1678, Cyprus; [2]Department of Chemistry, Duke University, Durham, NC 27708 USA; [3]Department of Chemical Physics, Weizmann Institute, Rehovot 76100, Israel; [4]School of Chemistry, Tel Aviv University, Tel Aviv 69978, Israel; [5]Department of Chemistry, University of Pittsburgh, Pittsburgh, PA 15260 USA

Electron transmission through chiral molecules induced by circularly polarized light can be very different for mirror-image structures, a peculiar fact given that the electronic energy spectra of the systems are identical. We propose that this asymmetry - as large as 10% for resonance transport- arises from different dynamical responses of the mirrored structures to coherent excitation. This behaviour is described in the context of a general novel phenomenon, *current transfer*, transfer of charge with its momentum information, that is outlined in terms of a simple tight binding model. This analysis makes it possible to account for the observed asymmetry in and off resonance, to characterize its dependence on the length of the molecular bridge and to examine effects of dephasing processes.

===

Recent experiments report on control of molecular phenomena with polarization-shaped light pulses [1,2,3]. Indeed, theoretical studies suggest that circularly-polarized light can induce molecular circular electronic currents that can be considerably larger than molecular ring currents induced by static magnetic fields.[4,5].

Two of us and coworkers[6,7] have shown that the relative yield of electron transfer (ET) induced by circularly polarized light through helical molecular bridges depends on the relative handedness of the bridge and of the optical circular polarization, in spite of the indistinguishability of the underlying electronic energy spectra. Reversing direction of the circular polarization or the molecular handedness has similar effects, while the molecular handedness does not influence the transmission of electrons generated by unpolarized light. In both experiments a larger electron transfer yield is measured when the molecular handedness matches the circular polarization. Yet, the magnitude of the effect observed in the resonance transmission process started in Ref. [6] (5-10%) is considerably larger than that found in Ref. [7] (~0.1%), where transfer takes place through on off resonance helical bridge.

In this paper we advance a simple tight-binding model that accounts for both the yield asymmetries and the difference in the magnitude of the effects seen in the resonance and off-resonace transport situations, in the context of the more general phenomenon of *current transfer*. By current transfer we refer to charge transfer in which the transferred charge carrier maintains at least some of its linear and/or angular momentum. A recent example of current transfer in photoemission is provided by Ref. [8], where a biased linear momentum distribution created on a Cu (100) surface is observed in the angular distribution of the photoemitted current. Fig. 1. shows several tight-binding models for current transfer. In all cases, at issue is the question whether electron transfer between donor D and acceptor A is affected by and/or carries information about the initial electron momentum states in D. In Figs. 1a and 1b, these states correspond to linear and circular currents, respectively. If some of the current directionality is preserved during tranfer then, as shown in Fig. 1c, the indicated circular current in the donor would produce a helical current in the helical bridge, whose clockwise orientation implies motion towards the acceptor. An opposite donor circular current induces an anticlockwise bridge helical current that tends to move in the opposite direction, implying a lower probability to reach the acceptor. This intuitive picture is substantiated below, using the fact that the nearest-neighbor tight-binding model discussed below it is equivalent to the linear model displayed in Fig. 1d. Realization of such (partial) conservation of linear or angular momentum in the charge-transfer process and its reflection in the electron transmission probability would account for the observations of Refs. [6,7], if we assume that the circularly polarized light excites a superposition of donor states with a finite angular momentum, similar to the ring currents discussed in Refs. [4,5].

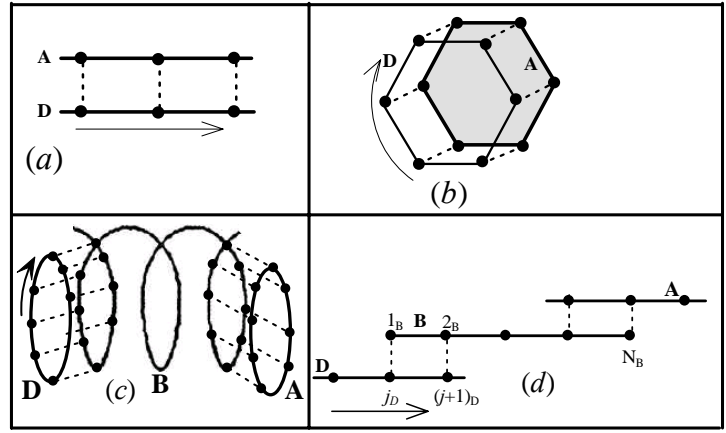

**Figure 1.** Shown are specific examples of current transfer from donor to acceptor (direct contact in a, b, transfer via intermediates in c and d). In the case of long or cyclic chains, the wave function amplitudes of the building blocks have well-defined phase relationships that produce the effects described here.

*Transport analysis.* We focus on the model of Fig. 1d, which describes the donor, bridge, and acceptor species as tight-binding chains.[9] The corresponding Hamiltonian is: $\hat{H} = \hat{H}_D + \hat{H}_A + \hat{H}_B + \hat{V}_{BA} + \hat{V}_{BD}$, where

$$H_K = \sum_{j_K \in K} \varepsilon_{j_K}^{(K)} |j_K\rangle\langle j_K| + \sum_{j_K \in K} V_{j_K, j_K+1}^{(K)} |j_K\rangle\langle j_K+1| \; ; \quad K = D, A, B \qquad (1)$$

and

$$V_{KK'} = \sum_{j_K \in K, j_{K'} \in K'} V_{j_K, j_{K'}}^{(K,K')} |j_K\rangle\langle j_{K'}| \; ; \quad (K, K') = (D, B) \text{ or } (B, A) \qquad (2)$$

$\hat{H}_D$, $\hat{H}_B$, and $\hat{H}_A$ are the Hamiltonians of the D, B and A moieties, respectively, while $\hat{V}_{DB}$ and $\hat{V}^{(BA)}$ are the D-B and B-A interactions. When the donor (say) is an $N_D$-member cyclic molecule, as in Fig. 1c, the periodicity is reflected by an additional cyclic boundary condition. A ring current in the donor is then represented by the complex quantum state

$$|M_D\rangle = \frac{1}{\sqrt{N_D}} \sum_{j_D=1}^{N_D} \exp[i(2\pi M_D(j_D - 1)/N_D)] |j_D\rangle. \qquad (3)$$

In the equivalent model of Fig. 1d, if coupling is assumed negligible among all but the nearest two sites of D and B, reversing the bridge handedness amounts to interchanging the coupling scheme from $j_D$-$1_B$ and $(j+1)_D$-$2_B$, shown in Fig 1d, to $(j+1)_D$-$1_B$ and $j_D$-$2_B$. Clearly, reversing the bridge handedness or the current direction $(M_D \rightarrow -M_D)$ has the same effect on the electron dynamics. Indeed, this symmetry is observed in the ET yields reported in Refs. [6,7].

We next examine the effect of current transfer on electron-transfer yield following the initial excitation of donor states $|\phi_{in}\rangle = |M_D\rangle$ and $|\phi_{in}^*\rangle = |-M_D\rangle$. For definiteness we assume that excited donor state is characterized by a finite lifetime $\hbar/\gamma_D$ and that the electron-transfer signal is associated with the decay of the acceptor state with rate $\hbar/\gamma_A$. These population relaxations are described by replacing $\varepsilon_{j_K}$ by $\varepsilon_{j_K} - i((1/2)\gamma_K)$ in Eq. (1) for the corresponding donor and acceptor sites, i.e., $\langle j_K | \hat{H} | j_K \rangle = \varepsilon_{j_K}^{(K)} - i(1/2)\gamma_K$, $K = D, A$.

Starting from states $|M_D\rangle$ and $|-M_D\rangle$, the yields for specific relaxation channels can then be calculated as follows. Starting from a given initial state $|\phi_{in}\rangle$, the probability that the acceptor state $|j_A\rangle$ is populated at time $t$, $P_{j_A,in}(t) = |\langle j_A | e^{-i\hat{H}t/\hbar} | \phi_{in} \rangle|^2$ ($\hat{H}$ is the (non-hermitian) Hamiltonian of Eq (1) with site-energies $\varepsilon_j - i((1/2)\gamma_j)$) can be computed in terms of the right and left eigenvectors, $|X_n^{(R)}\rangle$ and $\langle X_n^{(L)}|$ of $\hat{H}$, and the corresponding eigenvalues, $\mathcal{E}_n = E_n - i\Gamma_n/2$ ($\Gamma_n > 0$):

$$\langle j_A | e^{-i\hat{H}t/\hbar} | \phi_{in} \rangle = \sum_n R_{j_A,in}^{(n)} e^{-i\mathcal{E}_n t/\hbar}; \quad R_{j_A,in}^{(n)} = \langle j_A | X_n^{(R)} \rangle \langle X_n^{(L)} | \phi_{in} \rangle. \quad (4)$$

The yield of the irreversible flux out of the acceptor is then

$$Y(in) = \gamma_A \int_0^\infty dt \sum_{j_A} P_{j_A,in}(t) \quad (5)$$

The asymmetry associated with the excitation circular polarization or, equivalently, with the molecular bridge handedness may be quantified by the yields obtained from the initial states $|\phi_{in}\rangle = |M_D\rangle$ and $|\phi_{in}^*\rangle = |-M_D\rangle$

$$\mathcal{A} \equiv \frac{Y(M_D) - Y(-M_D)}{Y(M_D) + Y(-M_D)} \quad (6)$$

*Dephasing.* Current transfer as described above, is a coherent phenomenon, sensitive to environmental dephasing interactions. To investigate this effect we incorporate additional relaxation of coherences in the site representation of the Liouville equation for the system's density matrix $\rho$

$$i\hbar \frac{d}{dt}\rho_{j,l}(t) = \sum_k \left[ H_{j,k}\rho_{k,l}(t) - \rho_{j,k}(t)H_{kl} \right] - \left\{ i(\gamma_j/2 + \gamma_l/2) + i\gamma_{jl} \right\}\rho_{j,l}(t), \quad (7)$$

where, as above, the population relaxation rates $\gamma_j$ are non-zero only for donor and acceptor states. The probability $P_{j_A,in}(t) = \rho_{j_A,j_A}(t)$ needed in (5) is obtained from (7) using the initial condition $\rho(t=0) = |\phi_{in}\rangle\langle\phi_{in}|$.

*Model calculation.* We demonstrate the concept using the minimal model of Fig 2: a bridge (sites 3 to N-1) interacting with a donor (represented by two sites 1 and 2 and an acceptor site, N). All bridge site-energies are taken equal ($\varepsilon^{(B)}_{j_B} = \varepsilon_{br}$) and similarly for the bridge nearest neighbor couplings ($V^B_{j_B,j_B+1} = \beta_{br}$) and for the donor-bridge and acceptor-bridge couplings ($V^{(D,B)}_{1,3} = V^{(D,B)}_{2,4} = V^{(B,A)}_{N-1,N} \equiv V$. The complex energies of the donor sites (1 and 2) and acceptor site (N) are taken to be $\varepsilon - i(1/2)\gamma_D$, $\varepsilon$, and $\varepsilon - i(1/2)\gamma_A$. The physics of the electron transmission asymmetry reported in Refs. [6,7] is captured by this model, by representing the opposite initial circular current on the donor by

$$|\phi_{in}\rangle = \frac{1}{\sqrt{2}}\left(|1\rangle + e^{i\theta}|2\rangle\right); \quad |\phi^*_{in}\rangle = \frac{1}{\sqrt{2}}\left(|1\rangle + e^{-i\theta}|2\rangle\right), \quad (8)$$

and taking for the acceptor state $|j_A\rangle = |N\rangle$. Using Eqs. (5) and (6) this leads to

$$\int_0^\infty dt\, P_{N,in}(t) = \hbar \sum_{n=1}^N \frac{|R^{(n)}_{N,in}|^2}{2\Gamma_n} + 2\hbar \sum_{n>m}^N \mathrm{Im}\left\{ \frac{R^{(n)}_{N,in}\{R^{(m)}_{N,in}\}^*}{(E_n - E_m) - i(\Gamma_n + \Gamma_m)} \right\}, \quad (9)$$

where $\mathcal{E}_m = E_m - i\Gamma_m$ are the eigenenergies of the dissipative Hamiltonian, and $R^{(n)}_{fi,in} = \langle\phi_{fi}|X^{(R)}_n\rangle\langle X^{(L)}_n|\phi_{in}\rangle$. In the Liouville formalism we solve Eq. (7) using $\|\rho(t=0)\rangle\rangle = |\phi_{in}\rangle\langle\phi_{in}|$ or $|\phi^*_{in}\rangle\langle\phi^*_{in}|$ with $\gamma_{D(A)}/2 > 0$. In particular, coherence relaxation on the donor is accounted for by taking $\gamma_{12} > 0$. In either case we use Eqs. (5) and (6) to calculate the yield asymmetry.

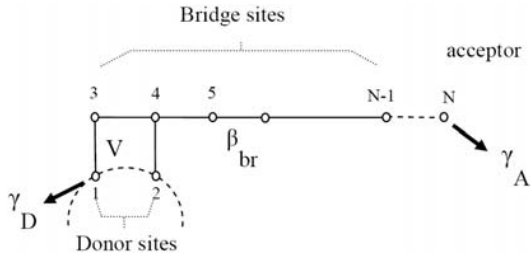

**Figure 2**. A minimal model demonstrating the effect of current transfer: The donor, containing sites 1 and 2 is coupled to the acceptor, site N, via the bridge, sites 3 to N. The population relaxation rates of the donor and acceptor sites are indicated with arrows, (rates $\gamma_D$ and $\gamma_A$). The initial states, Eq. (8), represent the initial current on the donor.

*Results.* Figs. 3 shows the asymmetry factor $\mathcal{A}$, Eq. (6), as a function of bridge length for resonant ($\varepsilon_{br} - \varepsilon = 0$) and non-resonant ($\varepsilon_{br} - \varepsilon = 3\text{eV}$) bridges, for different donor and acceptor lifetimes, ($\gamma_D = \gamma_A = 0.02, 0.2 \text{ eV}$) in the absence of dephasing. The yield asymmetry is seen to become independent of length for long bridges and to be about an order of magnitude larger in the resonant case. Furthermore, in both resonance and non-resonance cases $\mathcal{A}$ increases with decreasing donor and acceptor lifetimes, and more detailed studies show that the donor lifetime effect is dominant in this regard. Finally, for parameters in the range of those used here and for short enough donor lifetimes we find effects of the order seen in the experiments of Refs. [6] and [7] (~10% for resonant bridge, <1% in the off resonant case).

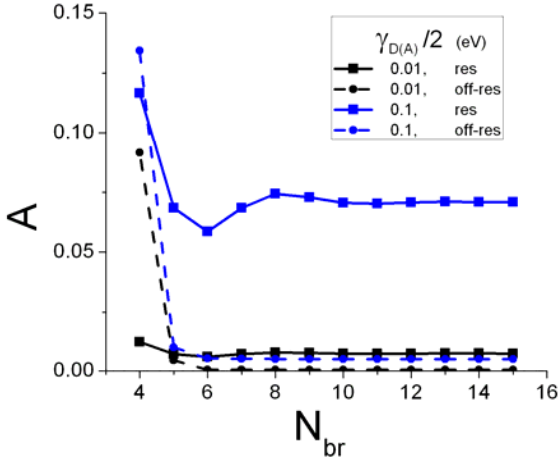

**Figure 3**. (color online) The yield asymmetry as a function of bridge length $N_{br} = N - 3$, for the model of Fig. 2, for resonant ($\varepsilon_{br} - \varepsilon = 0$, squares, solid lines) and off-resonant ($\varepsilon_{br} - \varepsilon = 3\text{eV}$, circles, dashed lines) bridges. Black and blue correspond to $\gamma_D = \gamma_A = 0.02 \text{ eV}$, and $\gamma_D = \gamma_A = 0.2 \text{ eV}$, respectively. Other parametrs used are $\beta_{br} = 0.5 \text{ eV}$, $V = 1.0 \text{ eV}$, $\theta = \pi/4$ and $\gamma_{i,j} = 0$ for $i \neq j$.

The effect of donor decoherence is shown in Fig. 4, which shows $\mathcal{A}$ vs $\gamma_{12}$ for resonant and non-resonant bridges of different lengths. A remarkable observation here is that although increasing decoherence eventually destroys the asymmetry, as expected, this happens only at unphysically large values of $\gamma_{12}$, in particular in the resonant bridge case. Similar results are obtained in the presence of decoherence on the bridge.

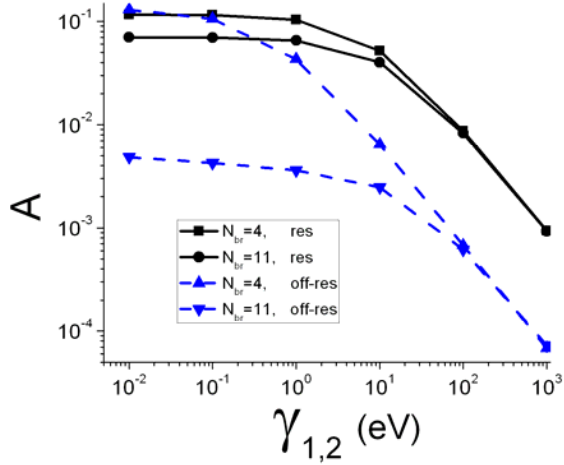

**Figure 4**. (color online) The yield asymmetry as a function of donor decoherence for the model of Fig. 2 for resonant and non-resonant bridges of different lengths. Parameters used (except $\gamma_{12}$) are the same as in Fig. 3 with $\gamma_D = \gamma_A = 0.2$ eV.

While this simple model cannot be expected to reproduce the specific experimental results of [6] and [7], we have found these distinctive and robust characteristics to characterize our model in a relatively large range of system parameters: 1) independence of bridge length for large molecular bridges, 2) asymmetry increases as donor lifetimes shortens and 3) asymmetry persists in the presence of decoherence. Furthermore, for a reasonable range of system parameters the calculated asymmetry is of the same order as seen experimentally under reversed handedness or circular polarization. Significantly, although this asymmetry is explained as a manifestation of a current transfer phenomenon resulting from coherent excitation, the resilience of this effect to decoherence rationalizes the observed behavior in condensed thermal environments.

It is important to note that our model does not rely inherently on the bridge chiral structure, but rather on the coherent nature of the donor excitation and on proximity effects (that, we propose, result from the chiral geometry) that determine the nature of the donor-bridge coupling. Such proximity effect are expected to be apparent in chiral molecules and in other nanostructures.

Two predictions highlighted in Figs 3 and 4 may be of particular value in probing molecular ET mechanisms: 1) the stronger yield asymmetry in the resonant regime and 2) the peaking of the yield asymmetries at short distances. These effects suggest that ET yield asymmetries may be used to distinguish between resonant and superexchange molecular charge-transfer mechanisms. Indeed, the transition between these regimes is of central interest in DNA electron transfer,[11,12] and direct strategies for mechanistic interrogation have proven elusive. A key experiment in DNA ET, therefore, would be to measure ET yield asymmetries for photoinduced ET with an intercalated ET active species excited with CPL. This experiment would be particularly informative if performed over a range of transport distances.

**Acknowledgments.**

This research was supported by the National Science Foundation (CHE-0718043, DNB; CHE-0415457 and CHE-0718755, DHW), the Israel Science Foundation, the German-Israel Foundation and the ERC (AN), the University of Cyprus, and the US-Israel BSF (AN, RN, DHW). We thank T. Fiebig, V. Mujica and M. Ratner for stimulating discussions.